\documentclass[aps,prd,a4paper,preprintnumbers,twocolumn,floatfix,showpacs,nofootinbib]{revtex4}

\usepackage{psfrag}
\usepackage{amssymb}
\usepackage{subfigure}
\usepackage{color}
\usepackage{mathrsfs}
\usepackage{graphicx}

\def\be{\begin{equation}}
\def\ee{\end{equation}}
\def\bea{\begin{eqnarray}}
\def\eea{\end{eqnarray}}

\begin{document}


\title{The Campanelli-Lousto  and veiled spacetimes}


\author{Luciano Vanzo}
\email[]{vanzo@science.unitn.it}
\affiliation{Dipartimento di Fisica, Universit\`a di Trento,
and INFN Gruppo Collegato di Trento, via Sommarive 14, 38100 Povo, Italy}

\author{Sergio Zerbini}
\email[]{zerbini@science.unitn.it}
\affiliation{Dipartimento di Fisica, Universit\`a di Trento,
and INFN Gruppo Collegato di Trento, via Sommarive 14, 38100 Povo, Italy}

\author{Valerio Faraoni}
\email[]{vfaraoni@ubishops.ca}
\affiliation{Physics Department and {\em STAR} Research Cluster, 
Bishop's University, 2600 College Street, 
Sherbrooke, Qu\'ebec, Canada J1M~1Z7}



\begin{abstract}
The Campanelli-Lousto solutions of Brans-Dicke theory, usually reported as 
black holes  are reconsidered and shown to describe, according to the values of a parameter, 
wormholes or naked singularities. The veiled 
Schwarzschild metric recently 
used as an example to discuss conformal frames and their equivalence corresponds 
to a special case of the CL metric. The conformal cousins 
of these solutions, and of the Riegert black hole solution of conformally invariant Weyl theory,  
are analysed, leading to a word of caution when interpreting physically spacetimes generated 
via conformal transformations from known seed solutions.
\end{abstract}


\maketitle




\section{\label{section1}Introduction}

There are relatively few exact solutions of alternative theories of gravity, although many 
such theories are currently studied with various motivations ranging from the possibility  
of using them to explain the current acceleration of the universe without dark energy, 
or for their properties in the early universe, 
as low-energy effective theories for quantum gravity, as emergent gravity theories, 
or just as toy models to understand which properties are, or are not, desirable in a theory 
of gravity (for recent reviews see  
 \cite{CliftonPadillaetc, SotiriouFaraoni, DeFeliceTsujikawa}). 

When getting to know a theory of gravity, it is important to understand its spherically symmetric
solutions and especially its black holes. The protoype alternative to Einstein's theory 
of General Relativity (GR) was Brans-Dicke theory \cite{BD}, 
later generalized to scalar-tensor gravity \cite{ST}. An early classification of 
spherical solutions of Brans-Dicke theory was given by Brans 
\cite{Bransspherical}.
 A common tool used in scalar-tensor and 
other theories of gravity is that of conformal transformations, which relate the physics 
in one conformal representation of the theory (``Jordan frame'') to another (``Einstein frame'').
Conformal transformations are useful to generate solutions of a theory from known seed 
solutions of another, but also to relate different 
solutions within the same theory. There has been a debate about the physical equivalence of conformal frames (see, {\em e.g.}, 
\cite{Flanagan, FaraoniNadeau} and the references therein) and recently the ``veiled Schwarzschild
spacetime'' has been used as an example for discussions of conformal frames 
\cite{DeruelleSasaki, ValerioAlex, AlexFirouzjaee}.    Among the known spherically 
symmetric and static solutions of Brans-Dicke theory are the Campanelli-Lousto spacetimes 
\cite{CampanelliLousto}, which
are usually reported as black holes or ``cold black holes'' 
(because they have zero surface gravity and temperature  
\cite{coldBHs}). It turns out that the 
veiled Schwarzschild spacetime is a special case of the Campanelli-Lousto metrics corresponding
to certain fixed values of the parameters. Given the 
widespread use of conformal transformations 
in gravity and in cosmology, one would like to understand better the veiled Schwarzschild metric
 and the more general Campanelli-Lousto class to which it belongs, as well as veiled black holes
in other theories of gravity.  Extra motivation is provided by the finding that generalized 
Brans-Dicke solutions describe asymptotically Lifschitz black holes in the Jordan (but not
in the Einstein) frame \cite{asympLifschitz}. 

In this paper we will first show (in Sec.~\ref{section2}) 
that the three-parameter Campanelli-Lousto class of 
solutions of Brans-Dicke theory does not describe black 
holes. It corresponds, instead, to wormholes or naked 
singularities, respectively, 
according to the value of one of the parameters.
We then identify, in Sec.~\ref{section3}, 
 the veiled Schwarzschild spacetime with a special case of 
the Campanelli-Lousto class and discuss its properties. A 
similar analysis is performed in Sec.~\ref{section4} 
for the veiled Riegert black hole, which is a solution of 
 Weyl's theory of gravity. It is found that caution is needed not to confuse 
Einstein frame metrics with scaling units of mass, length, and time with 
their versions with fixed units, and 
that special care must be taken when interpreting physically even straightforward 
mathematical results (Sec.~\ref{section5}). Sec.~\ref{section6} contains the conclusions.

\section{\label{section2}The Campanelli-Lousto solutions of Brans-Dicke theory}

 To begin with, we recall the Brans-Dicke field 
equations
\begin{eqnarray}
R_{\mu\nu}-\frac{1}{2} \, g_{\mu\nu} R &=& \frac{\omega}{\phi^2} \left( 
\nabla_{\mu}\phi \nabla_{\nu}\phi -\frac{1}{2}\, g_{\mu\nu} \nabla^{\alpha}\phi 
\nabla_{\alpha} \phi \right) \nonumber\\
&&\nonumber\\
&\, & +\frac{1}{\phi} \left( \nabla_{\mu} \nabla_{\nu}\phi -
\frac{1}{2}\, g_{\mu\nu} \Box \phi \right) \,, \label{BD1}\\
&&\Box \phi=0 
\label{BD2} \,.
\end{eqnarray}
The Campanelli-Lousto class of spherically symmetric solutions of Brans-Dicke theory 
\cite{CampanelliLousto} is given by\footnote{The notation 
differs slightly from that of 
Campanelli and Lousto  \cite{CampanelliLousto}: they denote 
$a$ with $-n$, $b$ with $m$, and $2\mu$ with $r_0$.} 
\be
ds^2=-V^{b+1}(r)dt^2+\frac{dr^2}{V^{a+1}(r)} +\frac{r^2}{V^{a}(r)} \, d\Omega_{(2)}^2 
\label{CLlineelement}
\ee
where 
\begin{eqnarray}
V(r) &=& 1-\frac{2\mu}{r} \,,\\
&&\nonumber\\
\phi(r) &=& \phi_0 V^{\frac{a-b}{2}}(r) \,,\label{CLscalarfield}
\end{eqnarray}
 $r>2\mu$, $d\Omega_{(2)}^2=d\theta^2+\sin^2 \theta \, d\varphi^2 $ 
is the line element on the unit 2-sphere, 
$\mu >0 ,  a $, and $b$ are parameters, and $\phi_0 $ is a positive constant. We use units in 
which the speed of light $c$ and Newton's constant $G$ are set equal to unity and we 
follow the conventions of Ref.~\cite{Wald}.

The Brans-Dicke coupling parameter is 
\be\label{BDomega}
\omega(a,b)=-2 \,\, \frac{\left( a^2+b^2-ab+a+b \right)}{\left( a-b\right)^2} \,.
\ee
Taking the trace of (\ref{BD1}) and making use of (\ref{BD2}), one has
\be
{R^{\mu}}_{\mu} =\frac{\omega}{\phi^2}\nabla^{\alpha}\phi \nabla_{\alpha} \phi\,.
\ee
Thus, on the Campanelli-Lousto solution, the Ricci scalar is
\be\label{Ricciscalar}
{R^{\alpha}}_{\alpha}= -2 \, \frac{ V^{a-1}(r)}{r^4} \, 
 \mu^2\left( a^2+b^2 -ab+a+b \right) \,.
\ee
The Campanelli-Lousto solutions were believed to be black holes and were presented as such in
\cite{CampanelliLousto}. However, they correspond to 
wormholes in a certain region of the 
parameter space and to naked singularity solutions in other regions. 
The Campanelli-Lousto class of solutions was rediscovered, with a different parametrization,
 by Agnese and La Camera \cite{room} who were apparently 
unaware of \cite{CampanelliLousto} and 
interpreted correctly these solutions as wormholes in the relevant range of parameters.  
In the following we provide an alternative analysis based  
on the introduction of the areal radius and the related 
apparent or trapping horizon.

As a first step, 
we note that  the areal radius\footnote{Note that the areal radius depends on the parameters 
$a$ and $\mu$, but is independent of $b$.} reads 
\be\label{CLarealradius}
R(r) = \frac{r}{V^{a/2}(r)} \,.
\ee
We must now distinguish two  cases, corresponding to the 
sign of the parameter $a$.

\subsection{The case $a \geq 0$}
\label{subsec1}

\begin{figure}[t]
\includegraphics[scale=0.34]{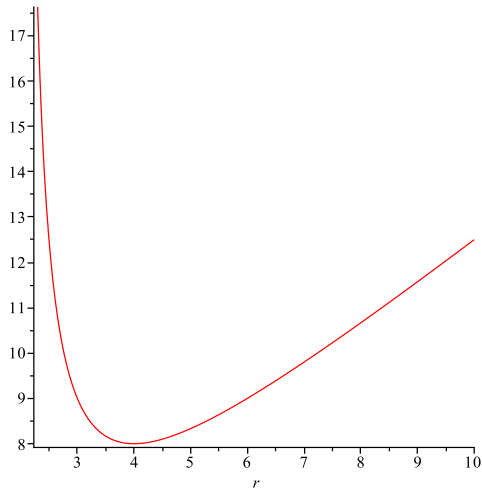}
\caption{\label{figA} the areal radius $R(r)$ as a function of $r$ (in units of $\mu$, for 
the parameter values  $\mu=1$ and $a=2$).} 
\end{figure}

\begin{figure}[t]
\includegraphics[scale=0.34]{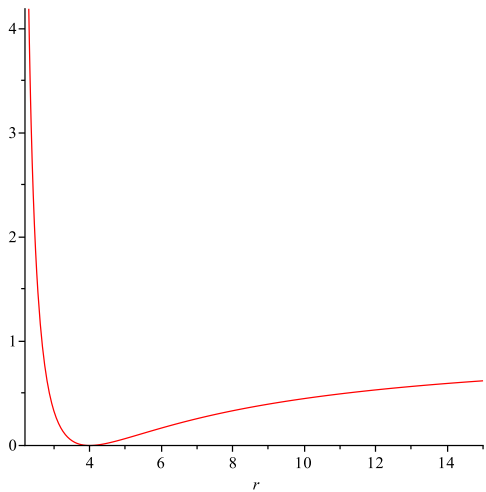}
\caption{\label{fig2} the metric coefficient $g^{RR}$ as a function of $r/\mu$ 
(for  the parameter values  $\mu=1$ and $a=2$).} 
\end{figure}

Let us study how the area of 2-spheres 
of symmetry behaves as $r$ varies.  We have 
\be
\frac{dR}{dr}= \frac{1}{V^{\frac{a+2}{2}}  } \left[ 1-\frac{ \left( 2+a \right) \mu}{r} \right]
\ee
and the areal radius (and consequently the area $4\pi R^2$ of 2-spheres which are orbits of
the isometry) decreases 
for $2\mu <r< r_{min} \equiv (2+a)\mu$, has a minimum 
\be
R_{min}\equiv R( r_{min})= (2+a) a^{-a/2} \mu
\ee
at $r_{min}$, and increases for  $r>r_{min}$ (fig.~\ref{figA}). 
By using the relation between differentials 
\be
dr=\frac{ V^{\frac{a+2}{2}}(r)}{1-(2+a)\mu/r} \, dR \,,
\ee
it is straightforward to rewrite the Campanelli-Lousto line element (\ref{CLlineelement})
 as
\be
ds^2=-V^{b+1}(r) dt^2 +\frac{V(r)}{\left[ 1-\frac{(2+a)\mu}{r} \right]^2} \, dR^2 +R^2 
d\Omega_{(2)}^2 \,.
\ee
The apparent (or trapping) horizons of a spherically symmetric spacetime
 are located by the equation $\nabla^cR\nabla_cR=0$, 
corresponding to $g^{RR}=0$ and $r=r_{min}$. Because of the exponent~2 in the denominator of 
the coefficient of $dR^2$, however, $g^{RR}$ does not change sign at its zero $r=r_{min}$ 
but has a double root there (see fig.~\ref{fig2}). Consider a bundle of radial outgoing null rays which start at 
$r<r_{min}$ and propagate outward. Their expansion $\theta_l$ is positive for $r<r_{min}$ 
(corresponding to the cross-sectional area of the bundle increasing as one moves along 
the bundle), it vanishes at $r=r_{min}$ (corresponding to a stationary cross-sectional area
of the bundle), and then it is positive again for $r>r_{min}$ (where this area begins 
increasing again). Therefore,  the apparent horizon at $r_{min}$ is not a black hole, but
rather  a wormhole apparent horizon. Thus, we answer the question posed in 
Ref.~\cite{amrita} asking whether wormholes supported 
solely by the Brans-Dicke scalar 
field are possible when the condition $\omega <-3/2$ is violated. 

\subsection{The case $a< 0$}
\label{subsec2}

\begin{figure}[t]
\includegraphics[scale=0.34]{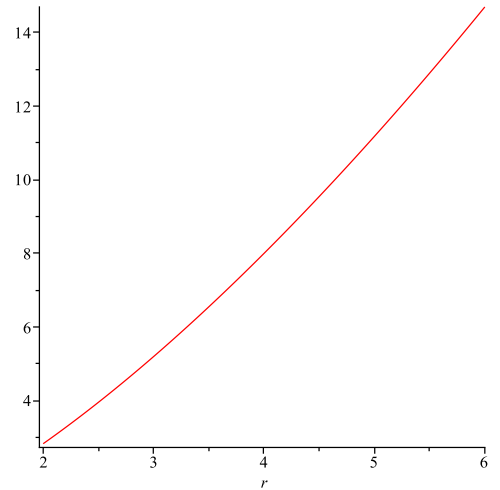}
\caption{\label{CampanelliLoustofig3} the areal radius $R$  as a function of $r$ 
(in units of $\mu$, for  the parameter values  $\mu=1$ and $a=-3$).} 
\end{figure}

If instead 
$a<0$, the areal radius becomes 
\be\label{questaR}
R(r)=rV^{\frac{|a|}{2}} (r)
\ee
and has a different shape. The minimum of the function (\ref{questaR}) 
occurs at $r_{\min}=(2+a)\mu <2\mu $ and $R(r)$ is always increasing in the 
relevant range $r>2\mu$, with $r\rightarrow +\infty $ corresponding to $R\rightarrow 
+\infty$ (see fig.~\ref{CampanelliLoustofig3}).  
In this case there is no apparent horizon.

The Ricci scalar is given by (\ref{Ricciscalar}) and, if $a<1$, diverges 
as $V \rightarrow 0$ when $r\rightarrow 2\mu^{+}$. The scalar field 
(\ref{CLscalarfield}) also vanishes there.  For $a<1$, therefore, 
the Campanelli-Lousto spacetimes contain a naked singularity at $R=0$ ({\em 
i.e.}, one that is not enclosed by a black hole horizon).  
Black holes, wormholes, and naked singularities can in principle be distinguished 
through the observation of their gravitational lensing 
effects  \cite{lensing}.

\subsection{The case $a=b$}

There remains to consider the special case $a=b$.  
The limit $b\rightarrow a $ corresponds to $\omega \rightarrow \infty$ in eq.~(\ref{BDomega}) 
and to the scalar field (\ref{CLscalarfield}) becoming  a constant: 
this is the limit to GR. Formally, in this limit 
the line element (\ref{CLlineelement}) reduces to
\be
ds^2=-V^{a+1}(r)dt^2+\frac{dr^2}{V^{a+1}(r)} +\frac{r^2}{V^{a}(r)} \, d\Omega_{(2)}^2 \,,
\ee
which is recognized as the Fisher-Janis-Newman-Winicour line element \cite{JanisNewmanWinicour}
\be\label{JNW}
ds^2=-V^{\nu}(r)dt^2+ V^{-\nu}(r) dr^2  + r^2 V^{1-\nu}(r)  \, d\Omega_{(2)}^2 
\ee
for $\nu=a+1$. This solution, rediscovered by various 
authors  \cite{Wyman}, is the most general 
static and spherically symmetric solution of the Einstein equations with zero 
cosmological constant and a massless scalar field  
\cite{Roberts}. 
However, by reducing to a constant, the scalar field of the Campanelli-Lousto solution
effectively disappears from the stress-energy tensor of the Brans-Dicke massless scalar 
field, which contains only first and second derivatives of $\phi$. 
Recall that the Brans-Dicke field 
equations (\ref{BD1}) and  (\ref{BD2}) reduce to the vacuum Einstein equations $R_{\mu\nu}=0$.
Setting the Ricci scalar~(\ref{Ricciscalar}) equal to zero when $b=a$  gives $a=-2, 0$.
The value  $a=b=0$ yields immediately the Schwarzschild solution with mass  $\mu$. 
For $a=b=-2$, as noted already in  \cite{CampanelliLousto}, 
the coordinate change 
$R= r-2\mu$ turns again the line element into the Schwarzschild one 
with mass parameter $m =-|\mu|$. It had to be so because, {\em in vacuo}, 
the Jebsen-Birkhoff theorem requires that  the only static 
spherically symmetric solution of $R_{\mu\nu}=0$ with zero cosmological constant be 
the Schwarzschild solution. This is also in agreement with a weak version of the 
Jebsen-Birkhoff theorem in scalar-tensor gravity  
\cite{VFBirkhoff}.

\section{\label{section3}The veiled Schwarzschild black hole}

In a recent paper, Deruelle and Sasaki discuss conformal transformations between different 
conformal frames, having in mind the Jordan frame and the Einstein frame used in scalar-tensor 
and $f(R)$ gravity, and argue the physical equivalence of these conformal frames. This paper 
should be placed in the context of  an ongoing 
 debate on the physical equivalence of conformal frames 
(especially in scalar-tensor gravity (see, {\em e.g.},  
\cite{Flanagan, FaraoniNadeau} and the references 
therein and in \cite{DeruelleSasaki}). Deruelle and Sasaki use as an example the ``veiled'' 
Schwarzschild black hole, {\em i.e.}, they take the 
Schwarzschild line element of GR 
\be\label{Schwarzschild}
ds^2=-\left(1-\frac{2M}{r} \right) dt^2 +\frac{dr^2}{1-\frac{2M}{r} } +r^2d\Omega_{(2)}^2 
\ee
as the Jordan frame metric and perform a conformal transformation to the Einstein frame metric 
$g_{\mu\nu} \longrightarrow \tilde{g}_{\mu\nu}=\Omega^2 g_{\mu\nu}$ with conformal factor 
$\Omega =1/\sqrt{ 1-2M/r}$. The result is the ``veiled'' Schwarzschild metric
\be \label{veiled}
d\tilde{s}^2=- dt^2 +\frac{dr^2}{\left( 1-\frac{2M}{r}\right)^2 } +
\frac{r^2}{  1-\frac{2M}{r} } \, d\Omega_{(2)}^2 \,.
\ee
The veiled Schwarzschild example is used also in  \cite{ValerioAlex} to show that the location of
an apparent horizon is not invariant under conformal transformations, and in 
\cite{AlexFirouzjaee} as an explicit example of an apparent (but not event) horizon to question 
standard beliefs about the thermodynamics of dynamical apparent horizons.

While, in general, conformally transforming a solution of a certain theory of gravity 
(including GR) does not produce another solution of that theory 
corresponding to the 
same form of matter (or to any reasonable 
mass-energy distribution), the point of  
\cite{DeruelleSasaki} is that the 
conformal transformation generates an equivalent representation 
 of the same physics provided that the
scaling of units in the Einstein frame 
length~$\sim \Omega$, time~$\sim \Omega$, and mass~$\sim \Omega^{-1}$ 
is taken into account, as explained long ago by Dicke 
\cite{Dicke}.  In other words, when developing a theory 
of gravity, it must be said how ordinary matter couples to 
gravity. In the Jordan frame matter is minimally coupled 
to the spacetime metric, the Einstein equivalence principle 
 holds, and units are fixed. In the Einstein 
frame, in which the gravitational part of the action 
assumes the Einstein-Hilbert form, matter is coupled to 
both the metric and the conformal factor $\Omega$ (or the 
scalar field $\phi$) and the units are no longer fixed 
(this property is true not only in the Einstein frame, but 
in all the ``veiled'' frames conformally related to the  
Jordan frame). The familiar results which 
follow from the Einstein equivalence principle do not hold 
in the ``veiled'' frames and either the notion of varying 
units or that of variable masses and couplings has to be 
introduced, expressing the fact that a minimally coupled 
theory has become a non-minimally coupled one. Then, a 
conformal transformation does not change the physics. 
However, one must be careful 
in using the  Einstein frame description since 
misunderstandings can occur, as explained below.

The veiled Schwarzschild metric is nothing but a special case of 
the Campanelli-Lousto solutions~(\ref{CLlineelement}) of Brans-Dicke 
theory corresponding to the parameter values $a=1, b=-1$, and $\mu=M$. 
According to the discussion of the previous section, this identification
 gives the areal radius 
\be
R(r)=\frac{r}{\sqrt{ 1-\frac{2M}{r}} }\,.
\ee
This function decreases between $2M <r < r_{min}=3M$, 
assumes the minimum value $R_{min}=R(3M)=3\sqrt{3}\, M$, and then increases for $r>3M$. As
 explained above for the general Campanelli-Lousto class of solutions, the 
metric~(\ref{veiled}), taken at face value,  describes a wormhole, not  a black hole. 
However, \cite{ValerioAlex} also notes that the 
quantity which is important to locate 
when considering conformal transformations is not the apparent horizon (which, contrary
to event horizons, changes location under a change of conformal frame). Instead, it is a 
new surface, characterized in \cite{ValerioAlex} in terms of an entropy 2-form, 
which is to be considered in place of the 
apparent horizon when conformal transformations are involved.  Indeed, it is noted in 
\cite{ValerioAlex} that there are no true trapping horizons (at $r=3M$ or elsewhere)
 in the veiled 
Schwarzschild metric (\ref{veiled}) and that the expansions of both ingoing and 
outgoing radial null geodesic congruences vanish at $r=3M$.
When the correct horizon defined through an entropy 2-form is taken into 
account, the metric (\ref{veiled}) is interpreted again as 
a black hole, not as a wormhole \cite{ValerioAlex}.  
This fact becomes more intuitive if one thinks that the apparent horizon is the place where the 
cross-sectional area ${\cal A}$ of a bundle of null rays becomes stationary and that, when the 
units of length are scaling in the Einstein frame, the units of area scale as $\Omega^2$ and
one must consider not ${\cal A}$ but the 
ratio of this quantity to its unit, 
$\sim {\cal A}/\Omega^2 $.
The ratio of the Einstein frame areal radius to its 
unit is $\sim R/\Omega =r$, which is trivially monotonic 
and does not describe a wormhole throat. The area of the 
black hole horizon expressed in varying units of area is 
\be
\frac{A}{\Omega^2} = 4\pi R_H^2 \left(1-\frac{2M}{r_H} 
\right) = 4\pi r_H^2=16\pi M^2 \,,
\ee
where $r_H=2M$.

What should we make of all this? If one forgets that the metric (\ref{veiled}) is 
obtained from the Schwarzschild one by means of a conformal transformation, one will naturally 
consider the apparent horizons of this spacetime instead of the new surface introduced in 
\cite{ValerioAlex} to characterize its nature. That is, 
one will correctly interpret the metric 
(\ref{veiled}) as a Campanelli-Lousto solution of Brans-Dicke theory representing a wormhole. 
However,  if the metric (\ref{veiled}) is instead seen as a conformally transformed 
Schwarzschild metric, which represents a genuine black hole (the prototypical one!) and 
one is aware that the apparent horizon of this metric is not 
the relevant quantity, but the relevant substitute of this concept is to 
be located using the prescription 
in \cite{ValerioAlex} which is conformally invariant, 
then~(\ref{veiled}) takes on a new meaning
 and is interpreted as a black hole instead of a wormhole. Note that both interpretations 
are correct {\em given the context to which they refer}, and that this
 context is very different in the two situations. If one does not know that 
(\ref{veiled}) comes from the
 conformal transformation of the Schwarzschild black hole outer region, the only admissible 
interpretation of (\ref{veiled}) is a wormhole spacetime. {\em Vice-versa}, if it is required that
(\ref{veiled}) is an Einstein frame metric arising from the conformal transformation 
of the Schwarzschild black hole, the interpretation is quite different:  units
are scaling in this frame, and the correct quantity to consider is not the surface where 
the expansions of null radial geodesic congruences vanish, but the surface introduced in 
\cite{ValerioAlex}.

To summarize, assigning a metric does not tell the whole story about a spacetime: if the 
metric arises from a conformal transformation of another metric, solution of a certain 
theory of gravity, this piece of information should be specified as an essential ingredient 
of the model because it determines the choice of quantities to be studied ({\em e.g.}, 
here, the apparent 
horizon versus the redefined horizon of  
\cite{ValerioAlex}), the whole context in which 
to study the metric  and, consequently, the 
physical interpretation of that spacetime metric.  This 
amounts to specifying that ordinary matter is non-minimally 
coupled to gravity in the Einstein frame.

One way to determine the mass of a black hole or spherical 
body operationally is by studying the deflection of light 
in 
the corresponding spacetime, a technique widely used to map 
the distribution of dark matter in galaxies and clusters. 
The famous Einstein formula for the deflection angle  of  
a light ray grazing the surface of a spherical body of mass 
$M$ and radius $R$ is $\delta 
\alpha=4M/R$ and is obtained in the weak-field 
approximation of the Schwarzschild solution. This 
deflection angle should also be obtained in the veiled 
Schwarzschild spacetime. {\em A priori} this is obvious 
because null geodesics are conformally invariant but, 
without this piece of information, it is not immediately 
clear that this is the case just by looking at the 
metric~(\ref{veiled}). Appendix~A shows how 
the calculation of the deflection angle and the empirical 
determination of the lens mass give the same result in the 
Schwarzschild and the veiled Schwarzschild spacetimes.

\section{\label{section4}Veiled conformal gravity  black holes}

The notion of veiled black hole may be generalized and, in order to have again a solution of 
the system one is dealing with, we consider a conformally invariant theory of gravity.  Thus, 
first we revisit Weyl gravity, namely a conformally invariant quadratic  gravity 
theory and its black hole solution found by Riegert and 
discussed by others \cite{r,m}. More 
precisely, we shall discuss the related topological black 
hole solutions  \cite{klem,seba}.
To begin with, we write down the action of the model, namely
\be
I =  \int_{\mathcal M} d^4 x \, C^2 \,, \label{d action}
\ee
where  $C^2 \equiv C_{\mu \nu \rho \delta} C^{\mu \nu \rho \delta}$ is the  
square of the Weyl  tensor. This conformally invariant gravity model  is 
very interesting and its phenomenology 
has been investigated in  \cite{m2} as a gravitational 
alternative to dark matter.
 
We are interested in spherically symmetric static solutions with a topological horizon, 
namely 
\be 
ds^2 = - W(r) dt^2 + \frac{dr^2}{W(r)} + r^2 d\Sigma_k^2\,, 
\label{d metric} 
\ee 
where 
\be 
d\Sigma^2_k=\frac{d\rho^2}{1-k\rho^2} + \rho^2 d\phi^2\,, 
\ee 
and the real  parameter $k$ can be $k=1$ (then the horizon manifold is the usual sphere 
$S_2$), $k=0$ (then 
the horizon manifold is the torus $T_2$), or $k=-1$ (in which case the horizon manifold is a compact 
hyperbolic manifold $Y_2$).

The topological Riegert solution can be written in the form  
\cite{seba} 
\be
W(r)=k+ 3c_0-\frac{c_0}{C}(2k+ 3c_0) r +b r^2-\frac{C}{r}\,,
\label{kfin}
\ee
where $c_0$, $C$, and $b$ are integration constants. The event horizon exists as soon as there 
is a positive real solution $r_H$ of the equation $W(r)=0$. For example, if $C >0$ and 
$b=\frac{1}{L^2}>0$, there exists always a positive root. Another simple case is $b>0$ and $c_0=0$, or $b=0$ and $k=0$. 

As already explained, the related topological ``veiled black hole'' may be obtained by a 
conformal transformation with $\Omega =1/\sqrt{W(r)}$, namely 
\be 
d \tilde{s}^2 = - dt^2 + \frac{dr^2}{W^2(r)} + \frac{r^2}{W(r)} \, d\Sigma_k^2 \,. 
\label{dd metric} 
\ee 
In order to analyse this metric, which is still a spherically symmetric static solution of Weyl 
conformal gravity, we recall the general Hayward formalism  
\cite{sean} and note that the 
spherically symmetric static spacetime can be written as 
\be 
d \tilde{s}^2= d \gamma^2+R^2(x)d\Sigma_k^2\,, 
\ee 
where the normal metric reads 
\be 
d \gamma^2 = \gamma_{ij}(x) \, dx^i dx^j=- dt^2 + \frac{dr^2}{W^2(r)} \,. \label{normal} 
\ee 
Thus, the new areal radius is 
\be 
R(r)=\frac{r}{\sqrt{W(r)}}\,. 
\ee 
The coordinate $r$ has a range such that $W(r) >0$. As a function of $r$, $R(r)$ diverges 
when $r \rightarrow r_H$. The trapping 
horizon is defined by 
\be 
\gamma^{ij}\partial_iR \, \partial_jR=W^2(r) \left(R'(r_0) \right)^2 =0\,, 
\ee 
and this reduces to the extremal condition $R'(r)=0$ for $R(r)$, namely 
\be 
W(r_0)=\frac{r_0}{2}\frac{d W_0}{d r}\,. 
\ee 
This condition defines a double ``trapping'' horizon at $r=r_0$, but without a trapped region 
since the two expansions of ingoing and outgoing radial null geodesic congruences 
are both vanishing at $r=r_0$ (see Appendix~B  
for further discussion). 
Making use of the Hayward invariant 
surface gravity, which can be defined in the presence of a generic trapping horizon 
\be 
\kappa_H=\frac{1}{2}\, \Box_{\gamma} R \Big\vert_H\,, \label{H} 
\ee 
one has in our static case 
\be 
\kappa_H=\frac{1}{2} \, W_0^2  R''_0 \,. \label{H1} 
\ee 
The second derivative of the areal radius at the apparent horizon is
\be
R''_0=\frac{1}{2W_0^{3/2}}\left( W'_0-r_0W''_0 \right)\,
\ee
Then, $R''_0 >0$ if and only if $\kappa_H >0$. Thus, $R$ has a local minimum and is locally 
 a wormhole since the lapse function is trivially constant. We may write
\be 
\kappa_H=\frac{1}{4} \, W_0^{1/2} \left(W'_0- r_0W''_0 \right) \,. \label{H2} 
\ee 
As an  example, for the veiled Schwarzschild black hole, 
where $W(r)=1-\frac{2M}{r}$, 
one has $\kappa_H=\frac{1}{6\sqrt{3}M}$ 
in agreement with \cite{AlexFirouzjaee}. 

A further local characterization may be achieved 
by introducing the Kodama vector, which is a natural generalization of the Killing vector 
$\partial_t$ to spherically symmetric static spacetimes. It is 
defined on the normal space and is trivially extended to the whole spacetime, namely 
\be 
K^i=\frac{\varepsilon^{i j} \, \partial_j R}{\sqrt{-\gamma}}\,.  
\ee 
In our case, one has 
\be 
K(r)=W(r)R'(r) \, \partial_t\,, 
\ee 
and at the double horizon $r=r_0$ where $R'_0=0$, it is $K_0=0$, 
which is a typical local property of wormholes.

Finally, we can investigate the asymptotic behaviour. When $r \rightarrow r_H$, the radius of the 
original black hole event horizon, the  veiled metric (\ref{dd metric}) reads
\be
d \tilde{s}^2 = - dt^2 + \frac{1}{\kappa_H^2 \,u^2}\left( d u^2+r_H^2 \, d\Sigma_k^2 \right)  
\label{ddd metric}
\ee 
where the new coordinate  $u$  is small, being  defined by 
\be
u^2=\frac{4}{W'_H}(r-r_H)\,,
\ee
and  $\kappa_H=W'_H/2$ is the  surface gravity of the original black hole. 
For  example for $k=0$, corresponding to an initial toroidal black hole, one has locally 
$ \mathbb{R} \times \mathbb{H}^3 $, $\mathbb{H}^3$ being the hyperbolic 3-dimensional space. 

With regard to the  other limit, it depends on the range of $r$. For example, if one assumes 
that $b>0$ and the original black hole is asymptotically anti-de Sitter then,  
when $r \rightarrow r_H$ (the original black hole event horizon radius), the veiled metric reads
\be
d \tilde{s}^2 = - dt^2 + \frac{1}{b^2} \, dv^2+ \frac{d\Sigma_k^2 }{b}  \,, 
\label{bddd metric}
\ee  
with the new coordinate defined by $v=1/r$, thus 
large $r$ corresponds to small 
values of $v$.

\section{\label{section5}Campanelli-Lousto spacetimes in GR are Fisher-Janis-Newman-Winicour}

If the veiled Schwarzschild metric (\ref{veiled}) is regarded as the conformal transform of 
the Schwarzschild black hole in an Einstein frame, with the information that all units are 
scaling and only ratios of quantities to the respective units are to be considered as physical 
(but not the quantities themselves), then (\ref{veiled}) is interpreted as describing a black 
hole. If instead the line element (\ref{veiled}) is taken without this information, 
{\em i.e.}, with fixed units, it is correctly interpreted as describing a 
wormhole spacetime. This situation, to which we arrived by 
contemplating the 
line element~(\ref{veiled}), has a counterpart in the 
general case of the Campanelli-Lousto 
solution (\ref{CLlineelement}), of which (\ref{veiled}) is a special case. That is, suppose 
that we perform the standard  conformal transformation to the Einstein frame of Brans-Dicke 
theory  $\left( g_{\mu\nu}, \phi \right)  \longrightarrow 
\left( \tilde{g}_{\mu\nu}, \tilde{\phi} \right) $ with 
\begin{eqnarray}
\tilde{g}_{\mu\nu} &=& \Omega^2 g_{\mu\nu}=\phi g_{\mu\nu} \,,\label{confo1}\\
&&\nonumber\\ 
\tilde{\phi}&=& \sqrt{ \frac{2\omega+3}{16\pi}} \, \ln \phi \,.\label{confo2}
\end{eqnarray}
Then the vacuum Brans-Dicke action 
\be\label{BDaction}
S_{BD}=\frac{1}{16\pi} \int d^4x \sqrt{-g} \left( \phi R -\frac{\omega}{\phi}\, g^{\mu\nu} 
\nabla_{\mu}\phi\nabla_{\nu}\phi \right)
\ee
is cast into the Einstein frame form
\be\label{GRaction}
S_{BD}=\int d^4x \sqrt{-\tilde{g}} 
\left(  \frac{\tilde{R}}{16\pi} -\frac{1}{2} \, \tilde{g}^{\mu\nu} \tilde{\nabla}_{\mu}
\tilde{\phi} \nabla_{\nu}\tilde{\phi} \right) \,,
\ee
where a tilde denotes Einstein frame quantities. If one forgets about the origin of this 
action and the scaling of units in the Einstein frame, one will interpret (\ref{GRaction}) 
as the action of GR with a minimally coupled and non self-interacting 
scalar field $\tilde{\phi}$ and fixed units. 
Then, the solutions of the vacuum Brans-Dicke theory will correspond to solutions of GR 
with this  scalar field. The Campanelli-Lousto class of solutions 
(\ref{CLlineelement}) generates a corresponding class of solutions of GR with
a free minimally coupled scalar field (and fixed units)  which  
coincides with the Fisher-Janis-Newman-Winicour  
\cite{JanisNewmanWinicour} class, as is shown 
below. The lesson is that, when the
information that varying units should be used in the Einstein frame is dropped, the 
Campanelli-Lousto wormholes do not correspond to black holes of Einstein theory.

Let us see how this situation occurs.  The Campanelli-Lousto line element conformally 
rescaled according to eq.~(\ref{confo1}) is
\be\label{43}
d\tilde{s}^2= -V^{\frac{a+b+2}{2}} (r) dt^2 +\frac{dr^2}{ V^{\frac{a+b+2}{2}}(r) }
+\frac{r^2}{ V^{\frac{a+b}{2}}(r)} \, d\Omega_{(2)}^2 \,,
\ee
where eq.~(\ref{CLscalarfield}) has been used and an irrelevant multiplicative constant 
has been dropped. According to eq.~(\ref{confo2}), the minimally coupled free scalar field 
sourcing this metric is
\be\label{CLfieldrescaled}
\tilde{\phi}=\frac{\pm (a-b)}{8\sqrt{\pi}} \sqrt{-(a+b)(a+b+4)} \, \ln V(r) 
\ee
if $b\neq a$. 
In order for this expression to be real, the argument of the square root must be non-negative, 
which corresponds to $\omega>-3/2$ and yields
\be\label{rangeofab}
-4 \leq a+b \leq 0 \,.
\ee
The solution depends only on the combination $\delta \equiv \frac{a+b}{2}$ and not on 
$a$ and $b$ separately and it is recognized to be of the Fisher-Janis-Newman-Winicour form 
(\ref{JNW}) with $\nu=\delta$, subject to the constraint $-2<\nu <0$.  
Since the Fisher-Janis-Newman-Winicour metric  is the most 
general static and spherically symmetric solution of the Einstein equations 
with zero cosmological constant and a massless scalar field 
as a source  \cite{Roberts}, 
the conformal transformation of the Campanelli-Lousto line element cannot give a  new 
solution of GR. The areal radius is
\be
R(r)= \frac{r}{V^{\frac{a+b}{4}} } 
\ee
(note that this expression depends on all the three parameters $a,b$, and $\mu$, contrary to 
the one in eq.~(\ref{CLarealradius})) and, since
\be
\frac{dR}{dr}= V^{-\frac{(a+b+4)}{4}}(r) \left[ 1-\frac{\mu}{r}\left( \frac{a+b+4}{2} \right) 
\right] \,,
\ee
formally the minimum of the function $R(r)$ is at $r_0=\left( \frac{a+b+4}{2} \right) \mu$. 
However, the range of the radial coordinate $r$ in the line element (\ref{CLlineelement}) is
$r>2\mu$ and $r_0>2\mu$  implies $ a+b \geq 0$, in contradiction with eq.~(\ref{rangeofab}). 
Therefore, it is always $r_0<2\mu$ and the areal radius $R(r)=r V^{\frac{|a+b|}{2}}(r) $ 
is always an  increasing function of $r$ for $r>2\mu$, with $\lim_{r\rightarrow 2\mu^{+}} 
R(r)=0$.

Now, using the relation between the differentials of the radial coordinates 
\be
dr= \frac{ r V^{\frac{a+b+4}{4}} (r) }{1-r_0/r} \, dR \,,
\ee
one obtains the line element 
\be \label{newCLsolutions}
d\tilde{s}^2= -V^{\frac{a+b+2}{2}} (r) dt^2 
+\frac{ V(r) }{\left( 1-\frac{r_0}{r}\right)^2}\, dR^2
+ R^2 \, d\Omega_{(2)}^2 \,.
\ee
Since we are now considering GR with fixed units (which is a 
very different context  from Brans-Dicke theory in the Einstein frame with varying units), 
the apparent horizons are located by the equation $g^{RR}=0$ or
\be
\frac{\left(1-r_0/r\right)^2}{V(r)}=0
\ee
which has no roots for $r>2\mu$ since $r_0<2\mu$. There are no apparent horizons and the 
GR-with-fixed-units  solutions generated by the Campanelli-Lousto metrics 
(\ref{CLlineelement}) only correspond to the second class of spacetimes discussed in 
Sec.~\ref{subsec2} and to a subclass of the Fisher-Janis-Newman-Winicour solutions.

The Ricci scalar is now
\be
{R^{\mu}}_{\mu}=\frac{\mu^2 V^{\frac{a+b-2}{2}}}{2r^4} (a-b)^2 \left[-(a+b)(a+b+4) \right]\,, 
\ee
and if $a+b<2$, which is the case here, diverges as $V\rightarrow 0$ when  
$r\rightarrow 2\mu$
 and $R\rightarrow 0$. Also  the  scalar field (\ref{CLfieldrescaled}) diverges in this limit
and the spacetime described by the line element  (\ref{newCLsolutions}) contains
 a naked singularity at $R=0$. It is well-known that the 
Fisher-Janis-Newman-Winicour solution
 exhibits  a naked singularity at $r=2\mu$. This result is 
consistent with that of  Ref.~ \cite{AgneseLaCamera} 
whose authors  
find that adding a scalar field to the exterior 
Reissner-Nordstr\"om or Kerr solutions shrinks the event 
horizon to a point. Ref.~\cite{DamourFarese} studies a 
more general metric in the Einstein frame and provides 
conditions for it to describe a wormhole. The line 
element~(\ref{43}) is less general than the spherically 
symmetric metric of \cite{DamourFarese} and does not admit 
wormholes. Our discussion of Sec.~\ref{section2} is an 
(expanded) Jordan frame version of the Einstein frame 
discussion of \cite{DamourFarese}.

\section{\label{section6}Conclusions}

We have studied known solutions of Brans-Dicke theory, Weyl theory, and GR
related by conformal mappings. First, the Campanelli-Lousto solutions of Brans-Dicke theory 
which are believed to describe black holes, are shown instead to correspond  
to wormhole spacetimes  
for positive values of the parameter $a$, and to spacetimes containing a naked singularity 
at $R=0$ when $a<1$.  Then, we realized that the veiled Schwarschild metric used as 
an example in the discussion of the physical equivalence of conformal frames coincides with the 
Campanelli-Lousto solution of Brans-Dicke theory (\ref{CLlineelement}) for 
 the parameter values  $a=1, b=-1$, and $\mu=M$. When the Campanelli-Lousto metrics are mapped 
to the Einstein frame and their conformal cousins are regarded as GR
 solutions ({\em i.e.}, with fixed units), they always generate 
a subclass of the Fisher-Janis-Newman-Winicour solution containing a naked singularity. 
The lack of a one-to-one correspondence between black holes in the Jordan and Einstein frames 
(in the absence of scaling units) was already noted in \cite{coldBHs, BloomfieldNandi}, 
although the difference
between scaling units and fixed units was not noted. 
The moral is that there is a big difference between the Einstein frame with varying units of time, 
length and mass (and of course, derived units scaling in the appropriate way) 
and GR with fixed units. The Campanelli-Lousto and veiled Schwarzschild 
spacetimes demonstrate this difference.  
We have then studied the veiled version of the Riegert black hole solution of Weyl gravity, 
which is a conformally invariant theory. Even in this situation, the conformal transformation 
of the Riegert black hole  generates, as a solution of the  same theory,  a wormhole 
without trapped regions.

Our discussion  induces a word of caution when using conformal transformations to an 
Einstein frame and  drawing physical interpretations of the mathematical results. 
In such situations, properties which seem obvious are not 
always true.

\section*{Appendix~A: Empirical 
determination of the  mass of unveiled and veiled black 
holes using light deflection}

Consider a photon starting at spatial infinity which, 
in 
the absence of a gravitational lens, would have as 
unperturbed path the $z$-axis and the unperturbed 
four-momentum 
$$
p^{\mu}_{(0)}= \left( 1,0,0, 1 \right)=\delta^{\mu}_0 +
\delta^{\mu}_3
$$
in Cartesian coordinates $\left( t,x,y,z \right)$. Let us 
introduce now a spherical lens of mass $M$ described by the 
Schwarzschild metric~(\ref{Schwarzschild})   and assume 
that lensing occurs in the weak-field regime. The photon 
four-momentum is now $ p^{\mu}=p^{\mu}_{(0)}+\delta p^{\mu} 
$ and satisfies the null geodesic equation
$$
\frac{dp^{\mu}}{d\lambda} +\Gamma^{\mu}_{\alpha\beta} 
p^{\alpha}p^{\beta} =0 
$$
which, to first order in the deflections, reduces to 
$$
\frac{d \left( \delta p^{\mu} \right)}{d\lambda} 
+\Gamma^{\mu}_{\alpha\beta} 
p^{\alpha}_{(0)} p^{\beta}_{(0)}  =0 \label{Delta} \,,
$$
where
$$
\Gamma^{\mu}_{\alpha\beta}=\frac{1}{2} \,\eta^{\mu\sigma}
\left( h_{\sigma\alpha \, , \beta}+ 
h_{\sigma\beta \, , 
\alpha}- h_{\alpha  \beta \,, \sigma} \right)
$$
and the metric is expressed as the Minkowski metric plus 
perturbations, $g_{\mu\nu}=\eta_{\mu\nu}+h_{\mu\nu}$. 
Integrating along the photon path between the source 
and the observer gives
$$
\delta p^{\mu}= -\int_S^O d\lambda \,  
\Gamma^{\mu}_{\alpha\beta} p^{\alpha}_{(0)} p^{\beta}_{(0)} 
\simeq 
-\int_S^O dz  \,
\Gamma^{\mu}_{\alpha\beta} p^{\alpha}_{(0)} p^{\beta}_{(0)} 
\,,
$$
where to first order it is legitimate to approximate the 
photon path with the unperturbed path (the $z$-axis) along 
which $\lambda \simeq z=t$.

In the veiled Schwarzschild spacetime~(\ref{veiled}) the 
null geodesic equation is still satisfied by the 
four-momentum $\tilde{p}^{\mu}$ of the photon,
$$
\frac{d \left( \delta \tilde{p}^{\mu} \right)}{d\lambda} 
+\tilde{\Gamma}^{\mu}_{\alpha\beta} 
\tilde{p}^{\alpha}_{(0)} \tilde{p}^{\beta}_{(0)}  =0  \,.
$$
Strictly speaking, a conformal transformation can change an 
affinely-parametrized geodesic into a non-affinely 
parametrized one, but it is possible to 
reparametrize the null geodesic using an affine parameter 
such that the $z$-axis coincides with the unperturbed 
photon path and $\tilde{p}^{\mu}_{(0)} = p^{\mu}_{(0)}=
\delta^{\mu}_0+\delta^{\mu}_3 $, and we assume that this 
has been done. Using the relation between the Christoffel 
symbols of conformally related spacetimes
$$
\tilde{\Gamma}^{\mu}_{\alpha\beta}= \Gamma^{\mu}_{\alpha\beta}
+\frac{1}{\Omega} \left( 
\delta^{\mu}_{\alpha} \, \partial_{\beta} \Omega +
\delta^{\mu}_{\beta} \, \partial_{\alpha} \Omega 
-g_{\alpha\beta} \partial^{\mu} \Omega  \right) \,,
$$
one obtains
\begin{eqnarray*}
&&\frac{d ( \delta \tilde{p}^{\mu}) }{d\lambda} 
+\Gamma^{\mu}_{\alpha\beta} \, 
p^{\alpha}_{(0)}p^{\beta}_{(0)}\nonumber\\
&&\nonumber\\
&&+\left[ 
\delta^{\mu}_{\alpha} \, \partial_{\beta} ( \ln \Omega) +
\delta^{\mu}_{\beta} \, \partial_{\alpha} ( \ln \Omega ) 
-g_{\alpha\beta} \partial^{\mu} ( \ln \Omega ) \right] 
p^{\alpha}_{(0)}p^{\beta}_{(0)}  \,.
\end{eqnarray*}
Integrating along the unperturbed photon path between 
source and observer yields
\begin{eqnarray*}
&&\delta \tilde{p}^{\mu} = \delta p^{\mu}\nonumber\\
&&\nonumber\\
&& - \int_S^O dz 
\left[ 
\delta^{\mu}_{\alpha} \, \partial_{\beta} ( \ln \Omega) +
\delta^{\mu}_{\beta} \, \partial_{\alpha} ( \ln \Omega ) 
-g_{\alpha\beta} \partial^{\mu} ( \ln \Omega ) \right] 
p^{\alpha}_{(0)}p^{\beta}_{(0)} 
\end{eqnarray*}
and using the fact that 
$$
p^{\alpha}_{(0)}p^{\beta}_{(0)} =
\delta^{\alpha}_0\delta^{\beta}_0 +
\delta^{\alpha}_0\delta^{\beta}_3+
\delta^{\alpha}_3\delta^{\beta}_0 +
\delta^{\alpha}_3 \delta^{\beta}_3 \,,
$$
it is 
$$
\delta \tilde{p}^{\mu} =\delta p^{\mu}
-4\delta^{\mu}_0 \left[ \ln \Omega \right]_S^O 
-4\delta^{\mu}_3 \left[ \ln \Omega \right]_S^O 
\,.
$$
Since the light source and the observer are in the 
asymptotic region, the terms in square brackets vanish and 
one obtains $\delta \tilde{p}^{\mu}=\delta p^{\mu}$, as 
expected from the independent knowledge that null 
geodesics are conformally invariant. This result 
then produces  the same deflection angle and the same 
operational determination of lens mass in the 
Schwarzschild metric and in the veiled Schwarzschild 
spacetime.

\section*{Appendix~B: The light sphere of 
the Reigert black hole}

Here we discuss a geometrical interpretation of the double trapping horizon of a 
veiled black hole-wormhole with regard to the so-called 
light sphere of the original unveiled 
black hole. 

We recall that, given a spherically symmetric black hole solution in the form
$$
ds^2 = - W(r) dt^2 + \frac{dr^2}{W(r)} + r^2 d\Omega_2^2 \,, 
$$
one can determine the associated light sphere by 
studying 
the equation of motion of classical 
relativistic massless particles. For a massless particle, 
it is well-known that the 
associated Lagrangian may be written in the form
$$
L=\frac{1}{2V} \, \frac{ds^2}{d \lambda^2}  \,,
$$
$V$ being the einbein and $\lambda$ a suitable evolution parameter. We may deparametrize this 
relativistic reparametrization-invariant system by making the choice $d \lambda=d\phi$, 
for angular variable. Then, choosing the other angular variable 
$\theta=\pi/2$,  eliminating the einbein $V$, and making use of two other constants of motion 
$k_0$ and $h$ associated with the conservation of energy and angular momentum, respectively,
a textbook approach provides the equation of motion for the trajectory as well as the first 
integral of motion
$$
\left( \frac{d r}{d \phi} \right)^2+r^2W(r)=\frac{k_0^2}{h^2} \, r^4 \,.
$$
As is well-known, it is convenient to make use of the 
Newton variable $u \equiv 1/r$. Then, 
the equation of motion of the light trajectory reduces to
$$
\frac{d^2  u}{d \phi^2}+u W(u)+\frac{u^2}{2}\frac{d W}{d u}=0\,.
$$
The light sphere is defined by $u=u_0$, describing a circular trajectory with constant radius. 
As a result, the radius is determined by
$$
u_0 W_0+\frac{u_0^2}{2} \, \frac{d W_0}{d u}=0\,.
$$  
In terms of the original radial coordinate $r$, one has
$$
W(r_0)=\frac{r_0}{2}\frac{d W_0}{d r}\,,
$$
which corresponds to the extremal property of the areal radius of the related  ``veiled 
black hole'' 
$$
d \tilde{s}^2 = - dt^2 + \frac{dr^2}{W^2(r)} + \frac{r^2}{W(r)} d\Omega_2^2  \,. 
$$
This property still holds true for a topological black hole. However, the light sphere may not 
exist. For example, let us consider the topological black 
holes in the presence of negative or 
positive cosmological constant $\Lambda$ \cite{vanzo, brill, mann}
where
$$
ds^2 = - V(r) dt^2 + \frac{dr^2}{V(r)} + r^2 d\Sigma_k^2\,, 
\label{vanzo}
$$
with
$$
V(r)=k-\frac{C}{r}-\frac{\Lambda}{3} \, r^2\,,
$$
and where $C$ is a mass parameter, which in the case $k=-1$ may be negative. The radius of 
the light sphere must satisfy
$$
r_0=\frac{3}{2 k} \, C\,.
$$
Note that this condition does not depend on $\Lambda$.  For example, for $k=0$ (toroidal 
black hole), there is no finite light sphere. For $k=1$, $r_0$ must be bigger than the event 
horizon, and for $\Lambda=0$, $C=2M$,  one obtains the  
well-known result for the Scharzschild 
solution.  For $k=-1$ and $\Lambda <0$, the condition may be satisfied for a restricted 
range of the mass parameter $C$.

\begin{acknowledgments}
LV and SZ would like to thank Sean Hayward and Giovanni Acquaviva for 
very useful discussions about wormholes. VF 
thanks Alex Nielsen for discussions on veiled metrics
 and Bishop's University and the Natural Sciences and 
Engineering  Research Council of Canada for financial suppport. 
\end{acknowledgments}



\end{document}